\documentclass[aps,prl,twocolumn,floatfix,showpacs,amsmath,amssymb]{revtex4}
\usepackage{graphicx}
\usepackage{amssymb}
\usepackage{dcolumn}
\usepackage{color} 
\begin{document}
\preprint{UCRL-JRNL-227256}
\title{Structure of $A=10-13$ nuclei with two- plus three-nucleon interactions
from chiral effective field theory}
\author{P. Navr\'atil,$^1$ V. G. Gueorguiev,$^1$ J. P. Vary,$^{1,2}$ 
W. E. Ormand,$^1$ and A. Nogga$^3$}
\affiliation{$^1$Lawrence Livermore National Laboratory, P.O. Box 808, L-414,
Livermore, CA  94551, USA \\
$^2$Department of Physics and Astronomy, Iowa State University, Ames,
Iowa 50011, USA \\
$^3$Forschungszentrum J\"{u}lich, Institut f\"{u}r Kernphysik (Theorie),
D-52425 J\"{u}lich, Germany}

\date{\today}

\begin{abstract}
Properties of finite nuclei are evaluated with two-nucleon (NN) and three-nucleon (NNN) 
interactions derived within chiral effective field theory (EFT).
The nuclear Hamiltonian is fixed by properties of the $A=2$ system, except
for two low-energy constants (LECs) that parameterize the short range NNN interaction.
We constrain those two LECs by a fit to the $A=3$
system binding energy and investigate sensitivity of $^4$He, $^6$Li, $^{10,11}$B 
and $^{12,13}$C properties to the variation of the constrained LECs. 
We identify a preferred choice that gives globally the best description. 
We demonstrate that the NNN interaction terms significantly improve 
the binding energies and spectra of mid-$p$-shell nuclei not just with 
the preferred choice of the LECs but even within a wide range of the constrained 
LECs. At the same time, we find that a very high quality description of these nuclei
requires further improvements to the chiral Hamiltonian.
\end{abstract}
\maketitle
%

The nuclear strong interaction has proven to be complicated 
and replete with ambiguities. However, chiral perturbation theory (ChPT) \cite{Weinberg} 
provides a promising bridge to the underlying theory, QCD, that could remove ambiguities. 
Beginning with the pionic or the nucleon-pion system \cite{bernard95} one works consistently 
with systems of increasing nucleon number \cite{ORK94,Bira,bedaque02a}. 
One makes use of spontaneous breaking of chiral symmetry to systematically 
expand the strong interaction in terms of a generic small momentum
and takes the explicit breaking of chiral symmetry into account by expanding 
in the pion mass. Thereby, the NN interaction, the NNN interaction 
and also $\pi$N scattering are related to each other. 
At the same time, the pion mass dependence of the interaction is known, which will
enable a connection to lattice QCD calculations in the future \cite{Beane06}.
Nuclear interactions are non-perturbative, because diagrams with purely nucleonic
intermediate states are enhanced \cite{Weinberg}. Therefore, the ChPT expansion 
is performed for the potential. Solving the Schr\"odinger equation for this potential 
then automatically sums diagrams with purely nucleonic intermediate 
states to all orders. Up to the fourth- or
next-to-next-to-next-to-leading order (N$^3$LO) of the ChPT, all the LECs can be 
determined by the $A=2$ data with the exception of two LECs that must be fitted
to properties of $A>2$ systems. The resulting Hamiltonian 
predicts all other nuclear properties, including those of heavier nuclei.
We demonstrate that this reductive program works 
to predict the properties on mid-$p-$shell nuclei with increasing accuracy when 
the NNN interaction is included.

\begin{figure}[t]
\centerline 
{\includegraphics[width=8.5cm]{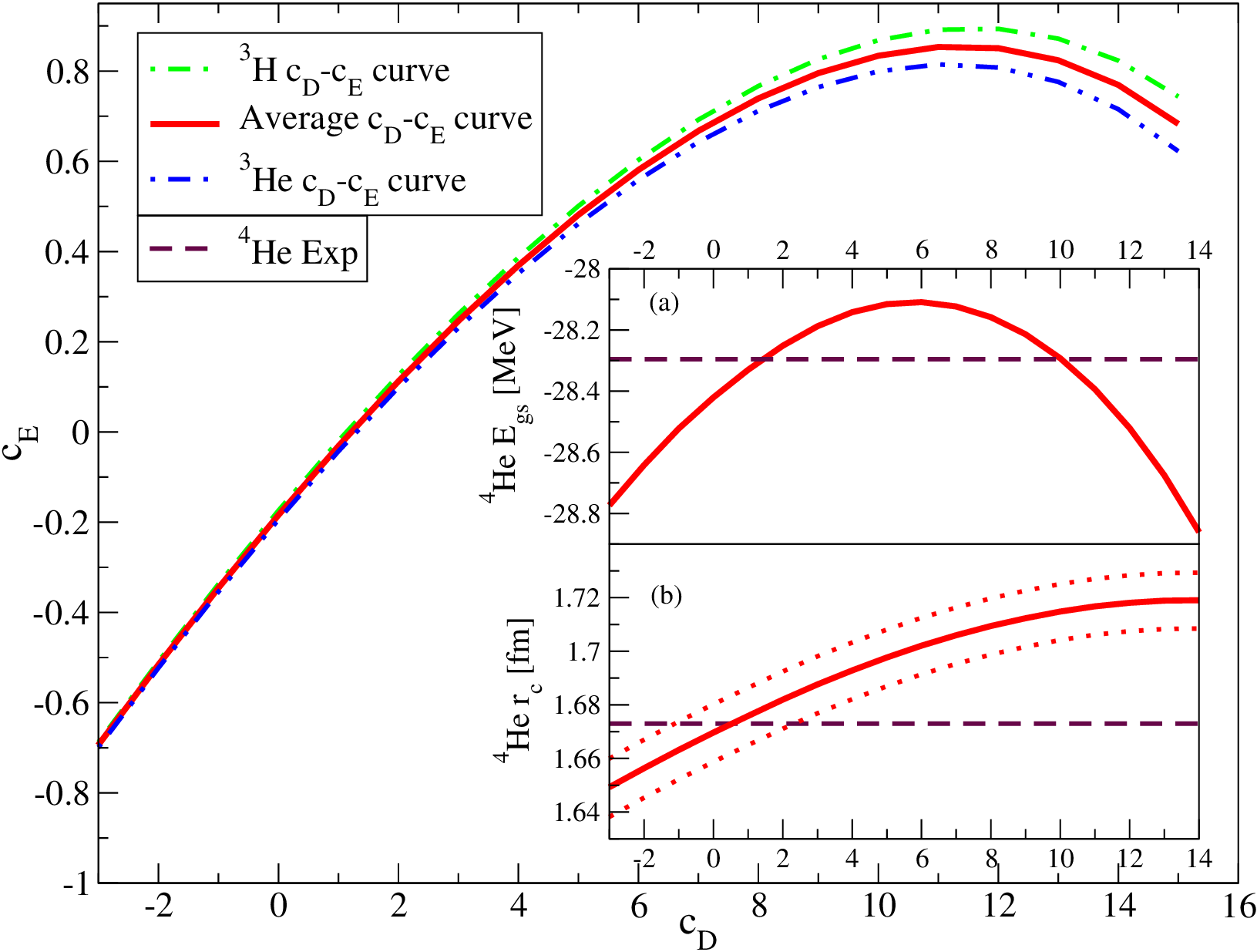}}
\caption{Relations between $c_D$ and $c_E$ for which  the 
binding energy of $^3$H ($8.482$ MeV) and  $^3$He ($7.718$ MeV) are reproduced. 
(a) $^4$He ground-state energy along the averaged curve. 
(b) $^4$He charge radius $r_c$ along the averaged curve. Dotted lines represent 
the $r_c$ uncertainty due to the uncertainties in the proton charge radius.}
\label{CDCE_curve}
\end{figure}

We adopt the potentials of ChPT at the orders presently available, the NN at N$^3$LO  
of Ref.~\cite{N3LO} and the NNN interaction at N$^2$LO \cite{EFT_V3b}.
Since the NN interaction is non-local, the {\it ab initio} no-core shell model 
(NCSM) \cite{NCSMC12,v3eff,NO03} is the only approach currently available 
to solve the resulting many-body Schr\"odinger equation for mid-$p$-shell nuclei.
In this paper, we use the NCSM to evaluate binding energies, spectra and other observables 
for $^6$Li, $^{10,11}$B and  $^{12,13}$C.  
We also present our results for the $s$-shell nuclei 
$^3$H, $^3$He and $^4$He. We use the $A=3$ binding energies to constrain  
the two unknown LECs of the NNN contact terms, $c_D$ and $c_E$ \cite{Nogga06}. 
We then investigate sensitivity of the $A>3$ nuclei properties to the variation 
of the constrained LECs.
Our approach differs in two aspects from the first NCSM application of the chiral 
NN+NNN interactions in Ref.~\cite{Nogga06} which presents 
a detailed investigation of $^7$Li.
First, we introduce a regulator depending on the momentum transfer in the NNN terms
which results in a local chiral NNN interaction. Second, we do not use exclusively 
the $^4$He binding energy as the second constraint on the $c_D$ and $c_E$ LECs.

\begin{figure}[t]
\begin{center}
{\includegraphics[width=8.7cm]{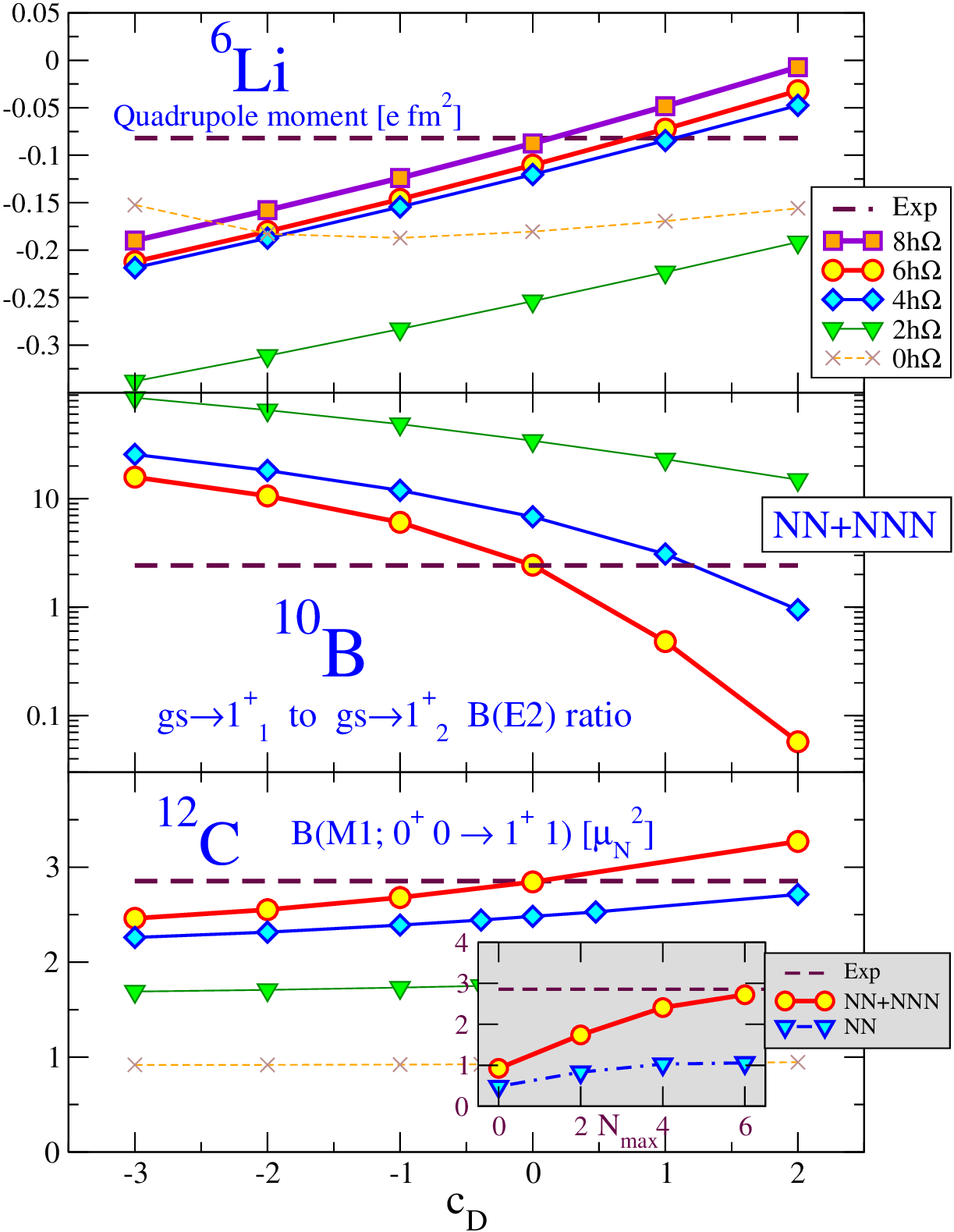}}
\caption{Dependence on the $c_D$ with the $c_E$ constrained by the $A=3$ 
binding energy fit for different basis sizes for: $^6$Li quadrupole moment, $^{10}$B 
B(E2;$3^+_1 0 \rightarrow 1^+_1 0$)/B(E2;$3^+_1 0 \rightarrow 1^+_2 0$) ratio, 
and the $^{12}$C B(M1;$0^+ 0\rightarrow 1^+ 1$). 
The HO frequency of $\hbar\Omega=13,14,15$ MeV was employed for 
$^6$Li, $^{10}$B, $^{12}$C, respectively. 
In the inset of the $^{12}$C figure, the convergence of the 
B(M1;$0^+ 0\rightarrow 1^+ 1$) is presented for calculations with 
(using $c_D=-1$) and without the NNN interaction.}
\label{cD_dep}
\end{center}
\end{figure}

The NCSM casts the diagonalization of the infinite dimensional many-body Hamiltonian matrix 
as a finite matrix problem in a harmonic oscillator (HO) basis with an equivalent 
``effective Hamiltonian" derived from the original Hamiltonian. The finite matrix problem 
is defined by $N_{\rm max}$, the maximum number of oscillator quanta shared by all nucleons 
above the lowest configuration. We solve for the effective Hamiltonian by approximating 
it as a 3-body interaction \cite{v3eff,NO03} based on our chosen chiral NN+NNN interaction.
With this ``cluster approximation", convergence is guaranteed with increasing $N_{\rm max}$.

It is important to note that our NCSM results through $A=4$ are fully converged in that 
they are independent of the cutoff, $N_{\rm max}$, and the HO energy, $\hbar\Omega$. 
For heavier systems, we characterize the approach to convergence by the dependence 
of results on $N_{\rm max}$ and $\hbar\Omega$.

\begin{figure}[t]
\centerline 
{\includegraphics[width=8.7cm,height=8cm]{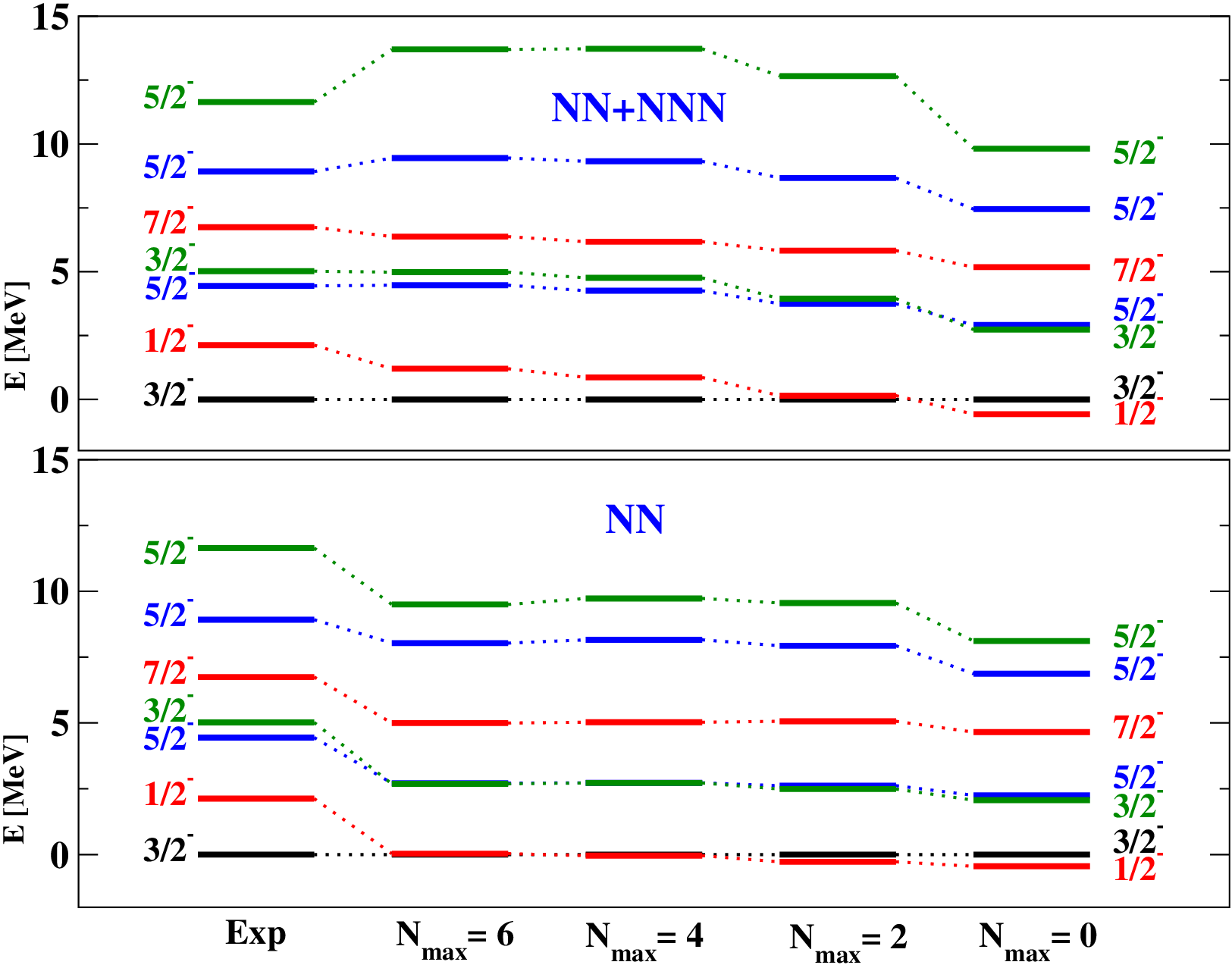}}
\caption{$^{11}$B excitation spectra as function of the basis space size $N_{\rm max}$ 
at $\hbar\Omega=15$ MeV and comparison with experiment. The isospin of the states 
depicted is $T$=1/2.}
\label{B11Spectral_convergence}
\end{figure}

Fig.~\ref{CDCE_curve} shows the trajectories of the two LECs $c_D-c_E$ that 
are determined from fitting the binding energies of the $A=3$ systems. 
Separate curves are shown for $^3$H and $^3$He fits, as well as their average.
There are two points where the binding of $^4$He is reproduced exactly.  
We observe, however, that in the whole investigated range of $c_D-c_E$, the calculated 
$^4$He  binding energy  is within a few hundred keV of experiment. 
Consequently, the determination of the LECs in this way is likely not very stringent.
We therefore investigate the sensitivity of the $p$-shell nuclear properties 
to the choice of the $c_D-c_E$ LECs. First, we maintain the $A=3$ binding energy 
constraint. Second, we limit ourselves to the $c_D$ values in the vicinity 
of the point $c_{D}\sim 1$ since the values close to the point 
$c_{D}\sim 10$ overestimate the $^4$He radius. 

While most of the $p$-shell nuclear properties, e.g. excitation spectra,
are not very sensitive to variations of $c_D$ in the range from -3 to +2 that 
we explored,
we identified several observables that do demonstrate strong dependence on $c_D$. 
In Fig.~\ref{cD_dep} we display the $^6$Li quadrupole moment that changes sign 
depending on the choice of $c_D$, the ratio of the B(E2) transitions from the $^{10}$B
ground state to the first and the second $1^+ 0$ state, and the $^{12}$C B(M1) transition
from the ground state to the $1^+ 1$ state. The B(M1) transition inset  
illustrates the importance of the NNN interaction in reproducing 
the experimental value \cite{Hayes03}. 
The $^{10}$B B(E2) ratio in particular changes by several orders of magnitude 
depending on the $c_D$ variation. This is due to the fact
that the structure of the two $1^+ 0$ states is exchanged depending on $c_D$. 
Using extrapolation, we can see that the best overall description is obtained 
around the $c_D\approx -1$. This observation
is also supported by excitation energy calculations as well as by calculations of
other transitions. We therefore select $c_D=-1$ and, from Fig.~\ref{CDCE_curve}, 
$c_E=-0.346$ for our further investigation.

We present in Fig. \ref{B11Spectral_convergence} the excitation spectra of $^{11}$B as 
a function of $N_{\rm max}$ for both the chiral NN+NNN, (top panel) as well as with 
the chiral NN interaction alone (bottom panel). In both cases, the convergence 
with increasing $N_{\rm max}$ is quite good especially for the lowest-lying states. 
Similar convergence rates are obtained for our other $p-$shell nuclei.

\begin{widetext}
\begin{figure}[htb]
\begin{center}
{\includegraphics[width=18cm]{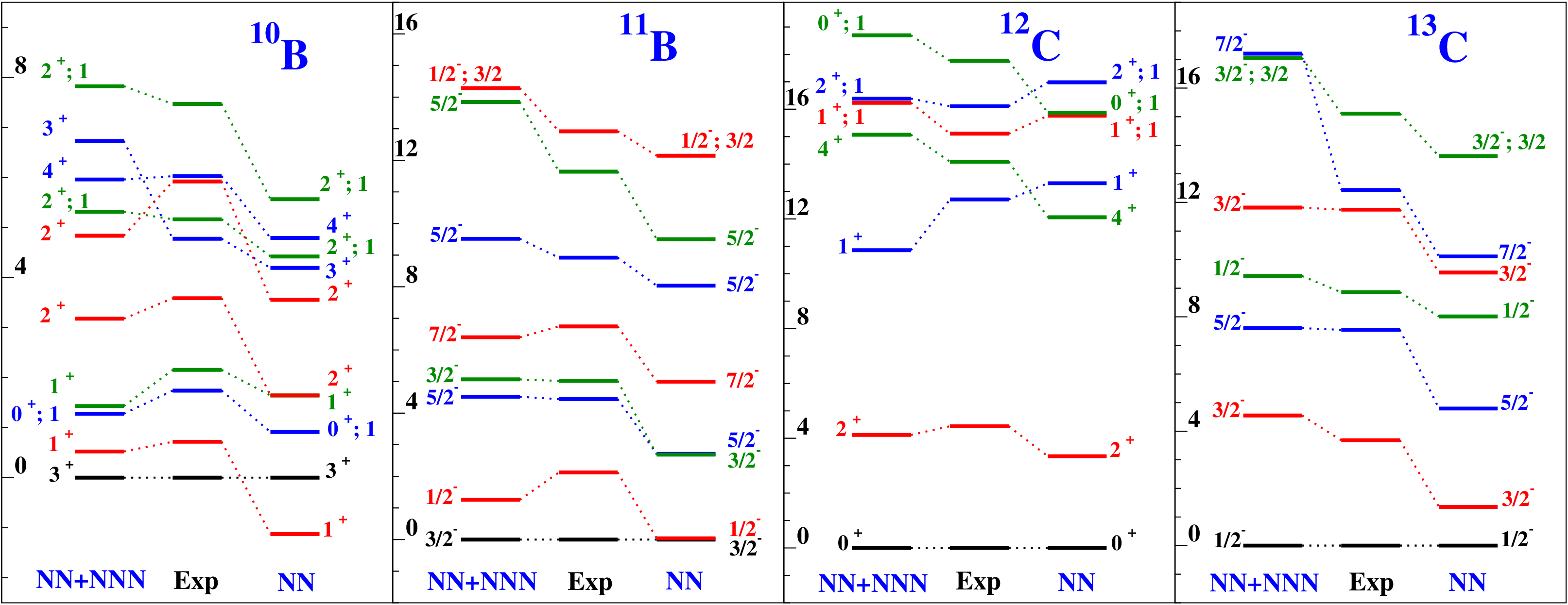}}
\caption{States dominated by $p$-shell configurations for $^{10}$B, $^{11}$B, $^{12}$C, 
and $^{13}$C calculated at $N_{\rm max}=6$ using  $\hbar\Omega=15$ MeV 
(14 MeV for $^{10}$B). Most of the eigenstates are isospin $T$=0 or 1/2, 
the isospin label is explicitly shown only for states with $T$=1 or 3/2.
The excitation energy scales are in MeV.}
\label{B10B11C12C13}
\end{center}
\end{figure}
\vspace{-0.7cm}
\end{widetext}

We display in Fig. \ref{B10B11C12C13} the natural parity excitation spectra of four nuclei 
in the middle of the $p-$shell with both the NN and the NN+NNN effective interactions 
from ChPT. The results shown are obtained 
in the largest basis spaces achieved to date for these nuclei with the NNN interactions, 
$N_{\rm max}=6$ ($6\hbar\Omega$). Overall, the NNN interaction contributes 
significantly to improve theory in comparison with experiment. 
This is especially well-demonstrated in the odd mass nuclei for the lowest few excited states. 
The celebrated case of the ground state spin of $^{10}$B and its sensitivity to the presence 
of the NNN interaction is clearly evident. There is an initial indication in these spectra 
that the chiral NNN interaction is ``over-correcting" the inadequacies of the NN interaction 
since, e.g. $1^+ 0$ and the $4^+ 0$ states in $^{12}$C
are not only interchanged but they are also spread apart 
more than the experimentally observed separation. 
While these results display a favorable 
trend with the addition of NNN interaction, there is room for additional improvement 
and we discuss the possibilities below. 

These results required substantial computer resources. 
A typical $N_{\rm max}=6$ spectrum shown in Fig. \ref{B10B11C12C13} 
and a set of additional experimental observables, 
takes 4 hours on 3500 processors of the LLNL's Thunder machine. 
We present only an illustrative subset of our results here.

\begin{table}
\caption{Selected properties of $^6$Li, $^{10,11}$B and $^{12,13}$C 
from experiment and theory. 
E2 transitions are in $e^2$ fm$^4$ and M1 transitions are in $\mu_N^2$.
The rms deviations of excited state energies are quoted for 
the states shown in Fig.~\protect\ref{B10B11C12C13}
whose spin-parity assignments are well established and that are known to be dominated 
by $p$-shell configurations. The total energy rms is for the 28 excited states 
from Fig.~\protect\ref{B10B11C12C13}.
Results were obtained in the basis spaces with $N_{\rm max}=6$ (8 for $^6$Li) 
and HO frequency $\hbar\Omega=15$ MeV (13 MeV for $^6$Li, 14 MeV for $^{10}$B). 
The experimental values are from 
Ref.~\protect\cite{AS88,AS90,AS91,T88,OZ88,Daito,Fujita04,Wang01}.}
\label{tabA10-13_m1bb}
{\center
\begin{tabular}{|c|c|c|c|}
\hline
  Nucleus/property & Expt.  & NN+NNN &  NN  \\ 
\hline
$^{6}{\rm Li}:~|E (1^+_1 0)|$ [MeV]  &  31.995  &  32.63  &  28.98  \\
$Q (1_1^+ 0)$ [$e$ fm$^2$]         & -0.082(2)&  -0.124 & -0.052  \\
$\mu (1_1^+ 0)$ [$\mu_N$]          & +0.822   &  +0.836 & +0.845  \\
$E_x (3^+_1 0)$ [MeV]              &  2.186   &   2.471 &  2.874  \\
B(E2;$3_1^+ 0 \rightarrow 1_1^+ 0$)& 10.69(84)&   3.685 &  4.512  \\   
B(E2;$2_1^+ 0 \rightarrow 1_1^+ 0$)&4.40(2.27)&   3.847 &  4.624  \\
B(M1;$0_1^+ 1 \rightarrow 1_1^+ 0$)& 15.43(32)&  15.038 & 15.089  \\
B(M1;$2_1^+ 1 \rightarrow 1_1^+ 0$)& 0.149(27)&   0.075 &  0.031  \\
\hline
$^{10}{\rm B}:~|E (3^+_1 0)|$ [MeV]  & 64.751      & 64.78   &  56.11  \\
$r_p$ [fm]                           &  2.30(12)   &  2.197  &  2.256  \\
$Q (3_1^+ 0)$ [$e$ fm$^2$]           & +8.472(56)  & +6.327  & +6.803  \\
$\mu (3_1^+ 0)$ [$\mu_N$]            & +1.801      & +1.837  & +1.853  \\
$ rms(Exp-Th) $     [MeV]            &    -        &  0.823  & 1.482   \\
B(E2;$1_1^+0 \rightarrow 3_1^+0$)    & 4.13(6)     & 3.047   & 4.380  \\
B(E2;$1_2^+0 \rightarrow 3_1^+0$)    & 1.71(0.26)  & 0.504   & 0.082  \\
B(GT;$3_1^+0 \rightarrow 2_1^+1$)    & 0.083(3)    & 0.070   & 0.102  \\
B(GT;$3_1^+0 \rightarrow 2_2^+1$)    & 0.95(13)    & 1.222   & 1.487  \\
\hline
$^{11}{\rm B}:~|E (\frac 32_1^- \frac 12 )|$ [MeV]   & 76.205      & 77.52   &  67.29 \\
$r_p(\frac 32_1^- \frac 12 )$ [fm]                   &  2.24(12)   & 2.127   &  2.196  \\
$Q (\frac 32_1^- \frac 12 )$ [$e$ fm$^2$]            & +4.065(26)  &+3.065   & +2.989  \\
$\mu (\frac 32_1^- \frac 12 )$ [$\mu_N$]             & +2.689      &+2.063   & +2.597  \\
$ rms(Exp-Th) $     [MeV]                            &  -          & 1.067   & 1.765   \\
B(E2;$\frac 32_1^- \frac 12 \rightarrow \frac 12_1^- \frac 12$) &   2.6(4)   & 1.476    & 0.750  \\
B(GT;$\frac 32_1^- \frac 12 \rightarrow \frac 32_1^- \frac 12$) &   0.345(8) &  0.235   & 0.663  \\
B(GT;$\frac 32_1^- \frac 12 \rightarrow \frac 12_1^- \frac 12$) &   0.440(22)&  0.461   & 0.841  \\
B(GT;$\frac 32_1^- \frac 12 \rightarrow \frac 52_1^- \frac 12$) &   0.526(27)&  0.526   & 0.394  \\
B(GT;$\frac 32_1^- \frac 12 \rightarrow \frac 32_2^- \frac 12$) &   0.525(27)&  0.762   & 0.574  \\
B(GT;$\frac 32_1^- \frac 12 \rightarrow \frac 52_2^- \frac 12$) &   0.461(23)&  0.829   & 0.236  \\
\hline
$^{12}{\rm C}:~|E (0^+_1 0)|$ [MeV] & 92.162     & 95.57   & 84.76  \\
$r_p$ [fm]                          & 2.35(2)    & 2.172   & 2.229  \\
$Q (2^+_1 0)$ [$e$ fm$^2$]          & +6(3)      & +4.318  & +4.931 \\
$ rms(Exp-Th) $     [MeV]           &    -       & 1.058   & 1.318  \\
B(E2;$2^+0 \rightarrow 0^+0$)    & 7.59(42)      & 4.252   & 5.483  \\
B(M1;$1^+0 \rightarrow 0^+0$)    & 0.0145(21)    & 0.006   & 0.003  \\
B(M1;$1^+1 \rightarrow 0^+0$)    & 0.951(20)     & 0.913   & 0.353  \\
B(E2;$2^+1 \rightarrow 0^+0$)    & 0.65(13)      & 0.451   & 0.301  \\
\hline
$^{13}{\rm C}:~|E (\frac 12_1^- \frac 12 )|$ [MeV]    & 97.108      &  103.23    &  90.31    \\
$r_p(\frac 12_1^- \frac 12 )$ [fm]                    &  2.29(3)    &  2.135     &  2.195     \\
$\mu (\frac 12_1^- \frac 12 )$ [$\mu_N$]              & +0.702      & +0.394     & +0.862   \\
$ rms(Exp-Th) $     [MeV]                             &    -        & 2.144      & 2.089  \\
B(E2;$\frac 32_1^- \frac 12 \rightarrow \frac 12_1^- \frac 12$) & 6.4(15)  & 2.659    & 4.584  \\
B(M1;$\frac 32_1^- \frac 12 \rightarrow \frac 12_1^- \frac 12$) & 0.70(7)  & 0.702    & 1.148  \\
B(GT;$\frac 12_1^- \frac 12 \rightarrow \frac 12_1^- \frac 12$) & 0.20(2)  &  0.095   & 0.328 \\
B(GT;$\frac 12_1^- \frac 12 \rightarrow \frac 32_1^- \frac 12$) & 1.06(8)  &  1.503   & 2.155  \\
B(GT;$\frac 12_1^- \frac 12 \rightarrow \frac 12_2^- \frac 12$) & 0.16(1)  &  0.733   & 0.263  \\
B(GT;$\frac 12_1^- \frac 12 \rightarrow \frac 32_2^- \frac 12$) & 0.39(3)  &  1.050   & 0.221  \\
B(GT;$\frac 12_1^- \frac 12 \rightarrow \frac 32_1^- \frac 32$) & 0.19(2)  &  0.400   & 0.151  \\
\hline
Total energy rms [MeV]     &   -   &1.314    & 1.671 \\
\hline
\end{tabular} }
\end{table}

Table \ref{tabA10-13_m1bb} contains selected experimental and theoretical results 
for $^6$Li and $A=10-13$. 
A total of 71 experimental data are summarized in this table including the excitation energies 
of 28 states encapsulated in the rms energy deviations. Note that the only case of an increase 
in the rms energy deviation with inclusion of NNN interaction is $^{13}$C and it arises due 
to the upward shift of the $\frac 72^-$ state seen in Fig. \ref{B10B11C12C13}, 
an indication of an overly strong correction arising from the chiral NNN interaction.
However, the experimental $\frac 72^-$ may have significant intruder components 
and is not well-matched with our state.

We demonstrated here that the chiral NNN interaction makes substantial contributions 
to improving the spectra and other observables. However, there is 
room for further improvement in comparison with experiment. 
We stress that we used a strength of the 2$\pi$-exchange piece of the NNN interaction,
which is consistent with the NN interaction that we employed. Since this strength 
is somewhat uncertain (see e.g. Ref.~\cite{Nogga06}), it will be
important to study the sensitivity of our results with respect to this strength.
Further on, it will be interesting to incorporate sub-leading NNN interactions and also
four-nucleon interactions, which are also order N$^3$LO \cite{Epelbaum06}.
Finally, we plan 
to extend the basis spaces to  $N_{\rm max}=8$ ($8\hbar\Omega$) for $A>6$ 
to further improve convergence.

Our overall conclusion is that these results provide major impetus for the full program 
of deriving the nucleon-nucleon interaction and its multi-nucleon partners in the consistent 
approach provided by chiral effective field theory. It is straightforward, but challenging, 
to extend this research thrust in the directions indicated. 
The favorable results to date and the need for addressing fundamental 
symmetries of strongly interacting systems with enhanced predictive power, 
firmly motivates this path.

This work was partly performed under the auspices of the
U. S. Department of Energy by the University of California, Lawrence
Livermore National Laboratory under contract No. W-7405-Eng-48. Support
from the LDRD contract No.~04--ERD--058 and from
U.S. DOE/SC/NP (Work Proposal No. SCW0498) 
and from the U. S. Department of Energy Grant DE-FG02-87ER40371 is acknowledged.
Numerical calculations have been performed at the LLNL LC and at the NERSC facilities, 
Berkely, and at the NIC, J\"ulich.

\end{document}